# The SiD Detector Concept

Andrew P. White[1]
(For the SiD Detector Concept [1])

1 – University of Texas at Arlington – Physics Department, 502 Yates Street, Arlington, Texas 76019, USA

The SiD detector concept is one of two currently accepted approaches to providing a detector for the future International Linear Collider. The SiD design philosophy is described and an overview of the concept given. Each major component of SiD is described, with emphasis on the critical R&D issues in each area. The context of the paper is evolution of the SiD design and the associated R&D towards the Detailed Baseline Design in late 2012.

## 1 Introduction

The SiD (or "Silicon Detector") is designed to be a compact, cost-contained detector for making precision measurements over a wide range of anticipated new phenomena at the future ILC. A quarter-section view of SiD is shown in Figure 1.

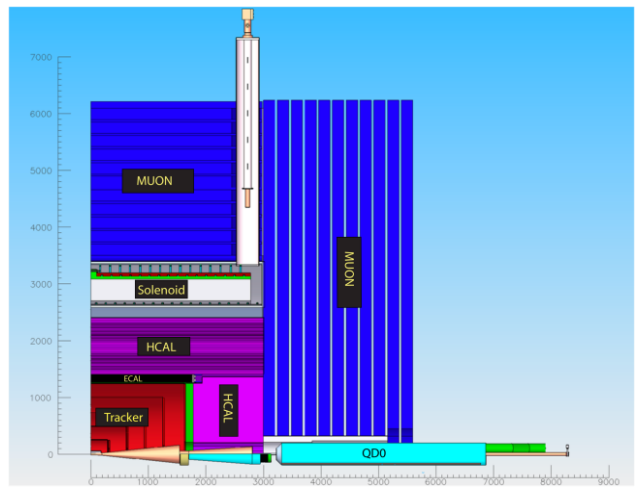

Figure 1. Quarter section view of the SiD detector.

At the heart of baseline SiD is a robust silicon vertexing and tracking system that offers excellent momentum resolution and is live for single bunch crossings. The electromagnetic calorimeter is also based on silicon sensors with 30 layers between tungsten absorber plates. The baseline hadronic calorimeter consists of RPC chambers between steel absorber plates. There is a very fine transverse subdivision of the calorimeter readout layers to support the use of the Particle Flow Algorithm approach to obtaining optimized jet energy resolution. The tracking and calorimeter systems are all located radially inside a 5 Tesla superconducting solenoid. The choice of the 5 Tesla field results from the need to compensate in terms of momentum resolution for the compact radial dimensions. External to the solenoid is an iron flux return and muon identifier, which acts as a component of the SiD self-shielding.

The focus for the SiD concept for the period through late 2012 is the Detailed Baseline Design (DBD). Last year the Letter of Intent was a major milestone for SiD and resulted in the validation of the SiD concept by the International Detector Advisory Group. Currently, there is ongoing R&D work in all subsystems as described in this paper. While there will be convergence on the design of the baselines for all major components to demonstrate the feasibility of the concept, it is clear that there will significant continued R&D beyond 2012.

## 2  Vertexing and Tracking

Figure 2. shows the layout of the vertex and tracking detectors.

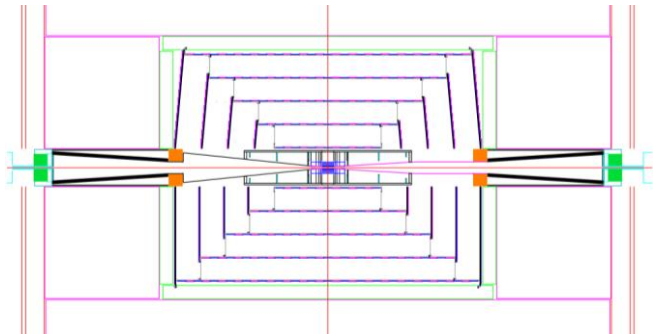

Figure 2. SiD vertex detector and silicon strip tracking system.

No ILC-ready vertex detector sensor yet exists. The critical R&D centers on developing one or more solutions for the sensors, a demonstrably stable and low mass mechanical support system, and pulsed power and cooling. Candidates for sensor technologies include the Chronopix, 3D, and DEPFET devices. Independent of the chosen technology, we expect the sensors to be glued on their edges to form cylinders, and the system to be gas cooled and power pulsed. As an example, Figure 3. shows the first Chronopix prototype under test, and the two sensor options for this prototype. A second prototype is expected in Fall 2010.

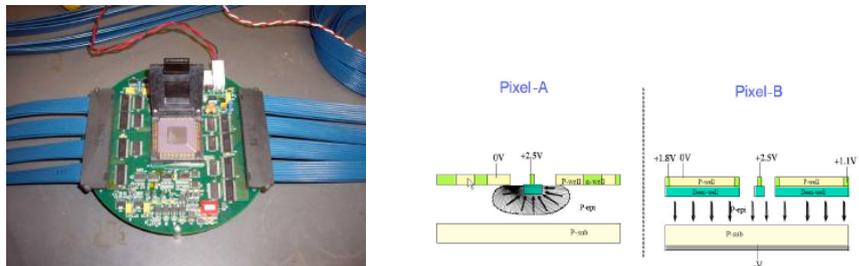

Figure 3.  Left – Chronopix sensor prototype under test. Right – sensor options.

SiD has an all-Silicon tracker consisting of 5 barrel (axial strip) and 4 disk (stereo strip) layers as shown in Figure 2. A specific design exists for "tiling" of the tracking layers with Silicon sensors and on-board KPiX readout chips (see below). Figure 4. shows a simulation of the tiles in the barrel tracking layers, together with a detail of the tiling arrangement.

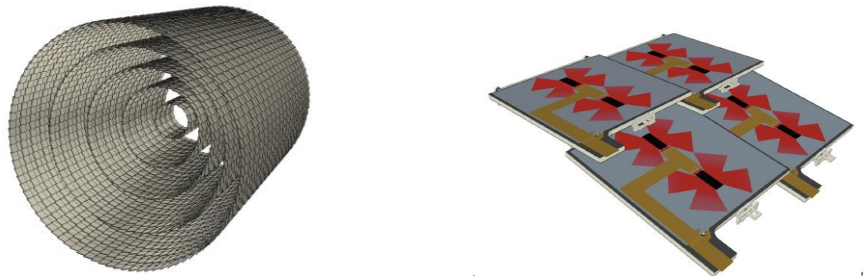

Figure 4. Left - simulation of the tiled sensor barrel layers. Right – tiled sensors.

The R&D priorities for the tracking system are the testing of a multi-sensor prototype, both with no magnetic field and in a 5Tesla field, understanding the optimal forward sensor configuration, and developing more detailed understanding of the mechanical stability and alignment system requirements.

Together with the vertex detector, the complete tracking system gives O(10) hits per track for reconstruction. Simulations and reconstruction have shown that the system has high tracking efficiency for tracks with transverse momenta above 0.2 GeV/c. SiD has also studied the tracking efficiency in the core of jets. The results are shown in Figure 5. along with the expected momentum resolution for a range of momenta and angles.

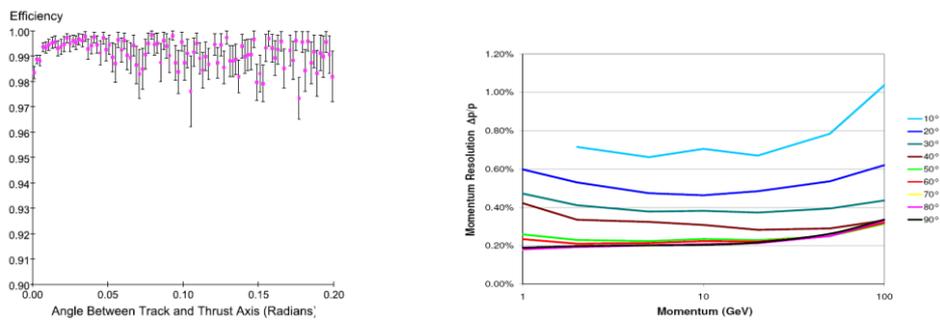

Figure 5. Left – Tracking system efficiency for jet cores. Right – Momentum resolution for a range of momenta and angles.

## 3  Electronics

A common KPiX chip is being developed to read out the tracker, calorimeters (but not the BeamCal), and muon systems. KPiX is a synchronous device (timed to the accelerator clock) and consists of a front-end amplifier with dynamic gain selection, an event trigger, a 4-deep pipeline feeding a 13-bit Wilkinson ADC, and calibration and leakage current subtraction features. The goal is to complete a 1024-channel version of the chip. A 512-channel version is now under test and extensive testing has previously been carried out for a 64-channel version. Figure 6. shows a complete test setup for a KPiX attached in this case to a digital hadron calorimeter anode board.

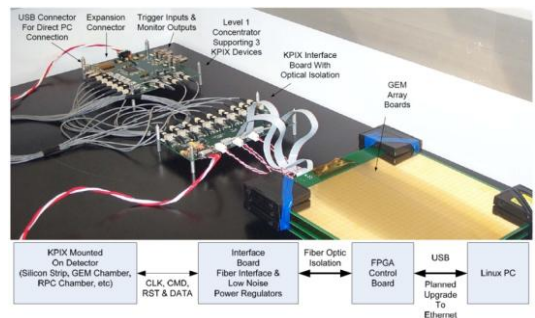

Figure 6. KPiX system components attached to a GEM anode board (bottom right).

In addition to the KPiX, other critical R&D items are the development of power distribution schemes for the vertex detector and tracker with DC-DC conversion or serial powering.

## 4  Calorimeters

The design of the SiD calorimeters derives from the adoption of the PFA approach to achieving jet energy resolution in the 3-4% range. The challenge is to follow precisely defined charged track trajectories from the tracker into the calorimeters and then associate each track with calorimeter energy deposits. For the electromagnetic calorimeter (ECal) this leads to a requirement for a small Moliere radius and fine transverse and longitudinal segmentation, for track following. For the hadronic calorimeter (HCal), similar segmentation requirements exist together with the need to optimize the HCal design within the radial space defined by the inner radius of the solenoid.

### 4.1  Electromagnetic calorimeter (ECal)

The baseline ECal is a Silicon-Tungsten device with thin sensors and electronics (KPiX) embedded between layers of Tungsten absorber. The critical R&D for the ECal centers on the demonstration of the operability of a fully integrated active layer within the 1.25mm gap between the absorber plates. Sufficient signal-to-noise, signal extraction, pulse powering, and adequate cooling must also be shown.  We are also considering the alternative MAPS (Monolithic Active Pixels) technology. The critical R&D for MAPS is to show the production of large sensors with sufficient yield.

Figure 7 shows the design of a baseline trapezoidal barrel ECal module and a detail of a hexagonal sensor layer. There are 30 layers with finer longitudinal segmentation in the front and coarser in the back of each module.

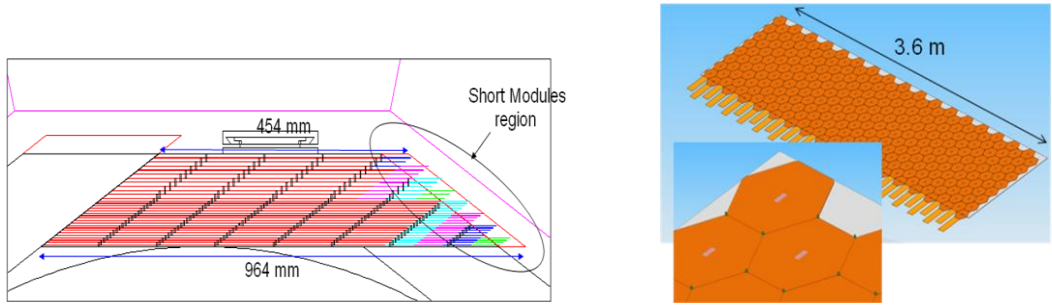

Figure 7. Left – Schematic of an ECal module. Right – Detail of the sensor arrangement.

The present R&D focus is the construction and testing of a 30-layer stack with one sensor per layer. This requires the development of the 1024-channel KPiX which is expected later in 2010. The stack will be exposed to an electron beam in 2011. Currently efforts are underway to finalize the technique(s) for connecting the flex readout cables and the KPiX chips to the silicon sensor layers. Figure 8 shows details of a possible gold stud bonding option.

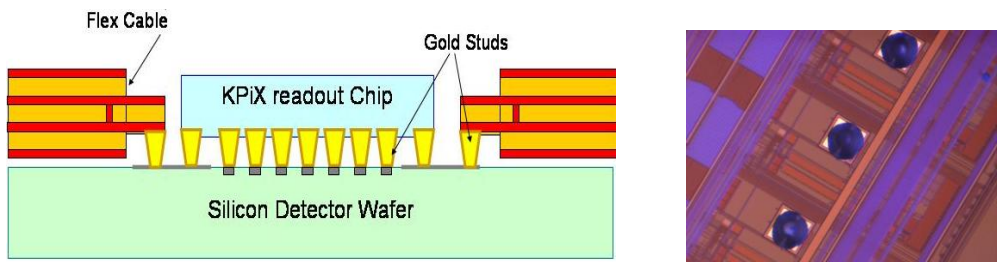

Figure 8. Left – Details of possible gold stud connections to a silicon layer. Right – gold stud bonding areas on a sensor.

### 4.2 Hadronic calorimeter

The HCal design and technology choices are still evolving. Figure 9 gives one possible barrel HCal design with non-projective cracks and some images from simulation of the effects of projective and non-projective cracks. For the HCal the most critical R&D priority is to demonstrate the feasibility of assembling a fully integrated, full size, active layer with a gap in the range 8-10mm between steel absorber plates. The technologies being considered for the active layers are RPC chambers (the current baseline choice), GEM chambers, Micromegas, and Scintillators with SiPM readout. A more radical alternative to all these is dual readout (scintillation and Cerenkov light) in a homogeneous crystal medium.

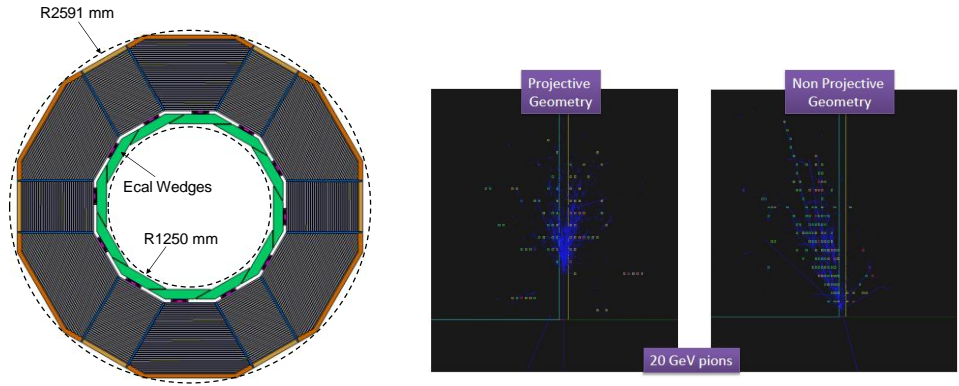

Figure 9. Left - Possible HCal design with non-projective cracks. Right – simulated single particle showers in projective and non-projective HCal geometries.

The baseline RPC active layer technology uses a double glass design. Shown in Figure 10. Following on from a successful "vertical slice test", that demonstrated the achievability of digital images and quantitative measurements of hadron showers, the present goal is the construction of a cubic meter prototype to contain hadron showers and explore the operability and stability of a large RPC system. Figure 10 also shows some details from the construction of square meter planes for the cubic meter stack.

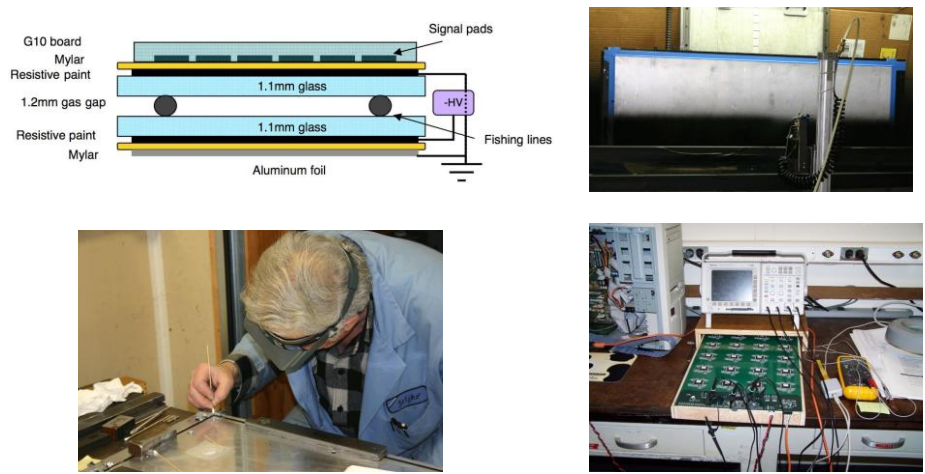

Figure 10. Left (top) – Schematic of double glass RPC. Left (bottom) – Assembling an RPC chamber. Right (top) – Applying the resistive coat to a glass panel. Right (bottom) – Testing a front-end board with on-board DCAL readout chips.

The micromegas HCal design is shown in Figure 11. Several small prototypes have been beam tested. The goal now is to produce meter square planes to test large-area performance.

Operating planes are expected in early 2011 using a new readout chip that is being developed for compatibility with the micromegas pulse shape. Figure 11 also shows a prototype meter square plane under construction.

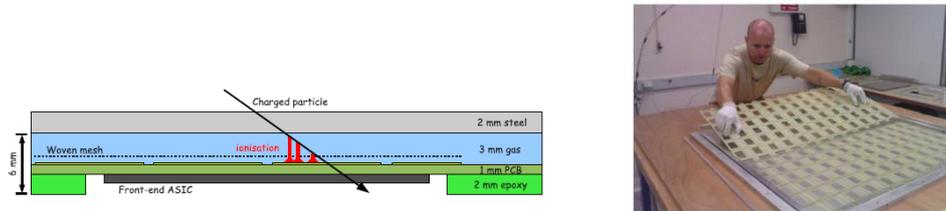

Figure 11. Left – Micromegas chamber schematic. Right – Assembly of large prototype plane.

The GEM (Gas Electron Multiplier) HCal design uses a double-GEM structure with on-board electronics as shown schematically in Figure 12. Extensive testing of a number of small prototypes has been carried out. An example result is given in Figure 12 with individual 1cm x 1cm anode pads showing the characteristic $^{55}$Fe double peak signal with an Argon/$CO_2$ mixture.

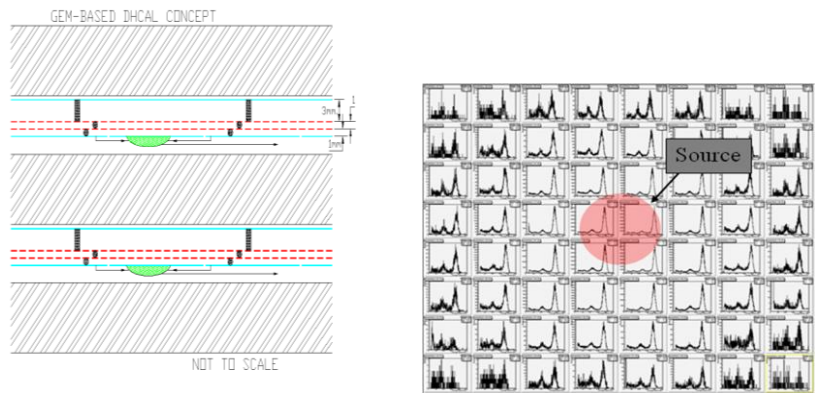

Figure 12. Left – The double-GEM HCal concept. Right – results of exposing an 8x8 anode 1cm x 1cm pad array to an $^{55}$Fe source.

The double-GEM HCal development plan foresees the inclusion of several meter square planes in the RPC cubic meter stack in 2011. Each meter square plan will consist of three 1m x 33cm GEM chambers. The necessary large foils are currently under fabrication in the CERN workshops.

For the Scintillator/SiPM HCal option, SiD is following the developments within the CALICE collaboration. Finally, the homogeneous dual-readout alternative awaits the development of suitable crystals, photodetectors, and associated readout electronics while preparing a demonstration of linearity and energy resolution for hadrons in a test beam.

### 4.3 Particle Flow Algorithm Development

SiD has been developing its own PFA. While satisfactory performance in terms of jet energy resolution and di-jet masses was achieved for the LOI, there was clearly scope for improvement. Algorithm development has focused on reduction of confusion from misasignments of energy deposits to the wrong track(s), and to handling of leakage through the outer dimensions of the calorimeter. Figure 13 shows an example of a misasignment when using a cone algorithm.

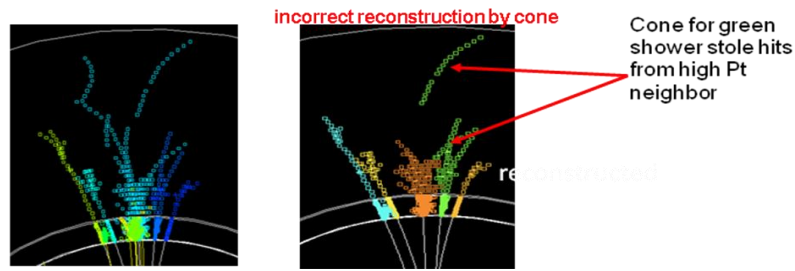

Figure 13 A detail of energy misasignment from SiD PFA development.

### 4.4 Muon system

Two technologies are being considered for the muon active layers: scintillator strips with SiPM readout, and Bakelite RPCs. Figure 14 shows some scintillator strips and the results of testing at at Fermilab.

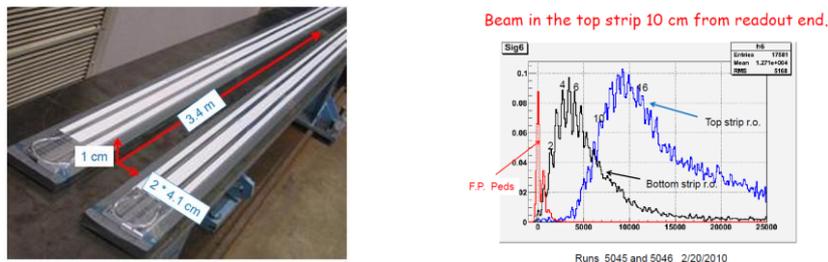

Figure 14. Left – Prototype muon scintillator strips. Right – signals from beam testing of strips.

The RPC option has been studied using chambers from the Babar endcap muon system. The results show good efficiency, but there have been signs of aging that is of concern. An aging study has shown the potential for a new type of Bakelite which has linseed oil impregnated – as for applications in BESIII and Daya Bay systems.

### 4.5 Machine Detector Interface

SiD has been engaged in studies of push-pull schemes (in a joint working group with ILD), vibrations, radiation simulations, and the magnetic field. As an example of work in this area Figure 15 (left) shows the present SiD scheme for rolling the detector in and out of the beam during the push-pull operations. The multi-roller solution allows a positioning accuracy of +/- 1 mm Figure 15 (right) shows a possible alternative push-pull solution, with both detectors on support legs.

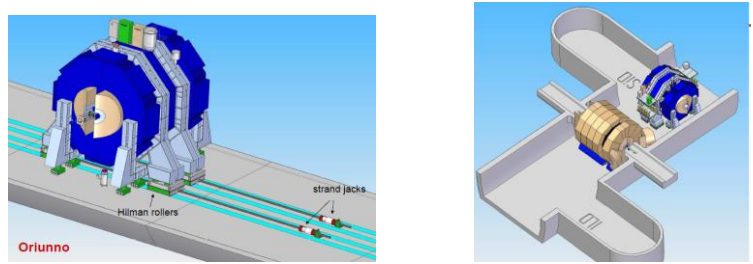

Figure 15. Left – The roller solution to push-pull for SiD. Right – A possible solution to joint push-pull for SiD and ILD, using support legs for both detectors.